\documentclass[12pt]{book}

\usepackage[dvips]{graphicx,color}
\usepackage{makeidx,tsukuba}

\makeauthorindex
\makeindex

\begin{document}

\BookTitle{\itshape The 28th International Cosmic Ray Conference}
\CopyRight{\copyright 2003 by Universal Academy Press, Inc.}
\pagenumbering{arabic}

\chapter{
Internal Structure of Ultra-High Energy Particles\\
with Lorentz Symmetry 
Violation at the Planck Scale}

\author{%
%
%
Luis Gonzalez-Mestres\\
{\it LAPP, CNRS-IN2P3, B.P. 110 , 74941 Annecy-le-Vieux Cedex}\\
}

\section*{Abstract}
Assuming the existence of a local vacuum rest frame (LVRF), and using suitable 
algebraic tranformations, the internal structure of ultra-high energy 
particles (UHEPs) is studied in the presence of Lorentz symmetry violation 
(LSV) at the Planck scale. Violations of the standard Lorentz contraction 
and time dilation formulae are made explicit. Dynamics in the rest frame of 
a UHEP is worked out and discussed. Phenomenological implications for ultra-high 
energy cosmic rays (UHECR), including 
possible violations of the Greisen-Zatsepin-Kuzmin GZK) cutoff, are 
studied for several LSV models.

\section{Models of Lorentz Symmetry Violation}

In the period 1995-97, we suggested several patterns of Lorentz symmetry
violation [4-7], all of them leading (for the "ordinary" sector of matter,
i.e. that with critical speed in vacuum equal or very close to the speed of light)
to deformed Lorentz symmetries (DLS) and to
deformed relativistic kinematics (DRK). Such LSV patterns 
were further discussed in subsequent 
papers, f.i. [8-13], especially quadratically deformed relativistic kinematics 
(QDRK) where for $k~a~\ll ~1$ ($k$ = wave vector, $a$ = fundamental length)
the particle energy $E$ can be written as:
\equation
E~~\simeq ~~ (2\pi )^{-1}~hc~a^{-1}~[(k~a)^2~-~\alpha ~(k~a)^4~
+~(2\pi ~a)^2~h^{-2}~m^2~c^2]^{1/2}
\endequation
\noindent
$h$ being the Planck constant and $c$ the speed of light. 
$\alpha $ is a model-dependent constant that may be in the range $0.1~-~0.01$ for
full-strength violation of Lorentz symmetry at the fundamental length scale,
and {\it m} the mass of the particle. For momentum $p~\gg ~mc$ , 
$E ~- ~ [p~c~+~m^2~c^3~(2~p)^{-1}]~= ~~\Delta E~\simeq ~ 
-~p~c~\alpha ~(k~a)^2/2$~.
It is assumed that the earth moves slowly with
respect to the absolute rest frame. For physical 
reasons developed in [14,15], we discarded 
linearly deformed relativistic kinematics (LDRK) which in the region
$k~a~\ll ~1$ is given by:
\equation
E~~ \simeq ~~ [(k~a)^2~-~\beta ~(k~a)^3~
+~(2\pi ~a)^2~h^{-2}~m^2~c^2]^{1/2}
\endequation
$\beta $ being a model-dependent constant. 
For $p~\gg ~mc$ , $\Delta E~\simeq ~ 
-~p~c~\beta ~(k~a)/2$ which is often invoked to explain
$TeV$ gamma data [1,2]. 

The analysis of these models, from the LSV 
point of view, was updated in [14-16] and has been further developed in [17,18]. 
LSV patterns based on mixing with superluminal sectors of matter are discussed
elsewhere in this Conference.
More generally, we shall mainly discuss here LSV patterns based on the appearance of a 
fundamental 
length scale which can be identified with the Planck scale but can also be interpreted
otherwise. However, the models we consider are different from the patterns proposed by 
Kirzhnits and Chechin [3,19,20] and by Sato and Tati [23], and should be compared with
them to make differences clear. 

The Kirzhnits-Chechin (KCh) model was based on a Finsler space. The relativistic
relation
$ p_{0}^2 ~ = ~ p^2~c^2 ~ + ~ m ^2~c^4$ (with $p^2~=~\Sigma_{i=1}^3~p_i^2$ , the $i$'s being space indexes) was replaced by:
$f ~ (p_{\mu}) ~ (p_{0}^2 ~ - ~ p^2~c^2) ~ = ~ m ^2~c^4$ ,
$\mu ~=~0,1,2,3$ standing for standard four-momentum indexes and
$f ~ (p_{\alpha})$ being a homogeneous positive function of the four-momenta of
zero degree. These authors used
$f ~ (p_{\mu }) ~ ~ = ~ ~ f ~ (\xi )$ where
$\xi ~ = ~ [p^2~c^2 ~ (p_{0}^2 ~ - ~ p^2~c^2)^{-1}]$ , 
$f ~ (0) ~ = ~ 1$ and $f$ was supposed to tend to some constant $f ~ (\infty )$ in
the range 0.01 - 0.1 as $\xi ~ \rightarrow ~ \infty $ . This amounts to a shift of
the effective squared mass by a factor 10 to 100 above some critical value of $\xi
$ . The dispersion relation for the photon was assumed to have no deformation, and
$f$ was taken to be $f ~ (p_{\mu }) ~ = ~ f ~ (\infty )$ for this particle. For a
massive particle, it was assumed that $f$ can be expanded as:
$f ~ (\xi ) ~ ~ \simeq ~ ~ 1 ~ - ~ \kappa ~ \xi ^2 ~ +~...$
leading, to a first sight, to models close to QDRK. For the proton, the term 
$\kappa
 ~ \xi ^2 $ becomes $\approx $ 1 at $E_p$ (proton energy)
$\approx $ $10^{20} ~ eV$ if $\kappa $ $
\approx $ $10^{-44}$ (the {\it ad hoc } choice to fit data).
In [18], we have worked out numerical examples and shown that actually the KCh pattern  
does not allow to build suitable models to explain the possible absence of GZK 
cutoff. But we also pointed out that, using an extension [17,18] of the 
Magueijo-Smolin operator formalism [21,22], the KCh pattern can be successfully modified
and unified with the QDRK approach in a larger class of models, including patterns with 
extra space-time dimensions.  

The Sato-Tati (ST) model was equally proposed as a solution to the UHECR puzzle. It implies the 
existence of a
preferred reference frame and the impossibility for hadronic matter to exist above a value of
the Lorentz factor $\simeq ~ 10^{11}$ with respect to this frame. Contrary to DRK, 
this model involves a very
strong dynamical assumption on the production of hadronic matter at ultra-high energy. 
Rather than with the structure of space-time, it seems 
to be concerned with the dynamical properties of vacuum in our Universe and with those 
of hadronic matter.
Even with a privileged LVRF, the ST model can incorporate
exact relativistic kinematics and have only a sharp dynamical threshold for the 
inhibition
of hadronic particle production. Furthermore, as discussed in [18], 
the suppression itself of the 
GZK cutoff in the ST model is unclear. 
However, the question of whether hadronic matter can exist above some critical value of $E/m$
in the LVRF is a fundamental one and certainly worth adressing. 
Some aspects of this problem are 
presently under study using the operator formalism developed in [17,18].

\section{Internal Structure and Space-Time Issues}

The idea of Doubly Special Relativity (see [1,2] and references therein) has
led to
the study of the possible formal equivalence between theories with Lorentz symmetry 
and theories with deformed Lorentz symmetry. This study led in turn to the suggestion  
by Magueijo and Smolin [21] to use an operator formalism to relate both kinds of theories.
In recent papers [17, 18], we further developed this idea and suggested new ways and 
potentialities. In particular, the operator formalism can be used to go to the rest frame of a 
UHEP and study the dynamics as seen by the particle. 
This would allow to examine, for the first time, 
fundamental dynamics such as it is seen by elementary particles in a
rest frame corresponding 
in the LVRF to a
momentum scale closer, in logarithmic scale, to Planck scale than to the 
electroweak scale
(e.g. for protons above $E~\approx ~10^{19}~eV$). There is no fundamental
reason for the laws of
Physics at these scales to look like we imagine them from laboratory formulations. 
There 
is even no compelling evidence that quantum mechanics is not violated for UHEPs 
together with special 
relativity. To study these crucial questions, 
the analysis of UHECR data 
may provide a powerful microscope directly focused on Planck scale dynamics. 

A simple illustration was provided long ago using simplified soliton models.
In a model using 
an analogy with the one-dimensional
Bravais lattice [7,9,10], it was shown that nonlocal effects at the
$\approx ~a$ scale
can change the internal structure of a relativistic object at distance scales
well above $a$    
(see also Gonzalez-Mestres, this Conference). With an example leading to QDRK, 
it was shown that if the typical size of a relativistic soliton is 
$\gamma ^{-1} ~\Delta $ , $\gamma $ being the effective 
Lorentz factor 
and $\Delta $ a characteristic
distance scale from soliton dynamics, 
the effective inverse squared Lorentz factor
$\gamma ^{-2}$ is corrected by a power series of $\xi $ ,
$\gamma ^{-2}~=~\gamma _R^{-2}~+~\gamma '~\xi ~+~...$,
$\gamma '$ being a constant of order 1 . Then, we expect the
departure
from standard relativity to play a leading role at
energies above that for which $\gamma _R^{-2}~\approx ~
\alpha ~(a~\gamma _R)^2~\Delta ^{-2}$ , i.e. above $E~\approx
~m~c^2~\alpha ~^{1/4}~(a~\Delta ^{-1})^{-1/2}$ . Taking the values
$\alpha ~\approx ~0.1$ , $m~\approx ~1~GeV/c^2$
and $\Delta ~\approx ~10^{-13}~cm$ , this
energy scale corresponds to $E$ above $\approx ~
2.10^{19}~eV$ for $a~\approx ~10^{-33}~cm$ .
Therefore, the internal structure of the UHEP changes drastically 
above this energy. That this is indeed the case can be checked
[17,18] using more recent techniques where an operator formalism
allows to go to the rest frame of the UHEP.

To roughly illustrate, without explicitly using operators, how 
a DRK boost technique can work, 
assume that in the LVRF a particle 
satisfies the QDRK:
\equation
p_{0}^2 ~ ~ = ~ ~ p^2 ~ ~ + ~ ~ m^2~ ~ - ~ ~ b ~ p^4
\endequation
\noindent
where $b$ is a constant, $b ~ m^2 ~ \ll ~ 1$ and we have taken $c~=~1$ . We can write 
the deformation term as: 
$b~ p^4 ~ ~ = ~ ~b~ (\pi _{\mu } ~ \pi ^{\mu })^2$
with 
$\pi _{\mu } ~ = ~ p_{\mu } ~ - ~ V^{-2} ~ (p_{\mu } ~ V^{\mu }) ~ V^
{\mu }$ , 
$V^{\mu }$ being a quadrivector with value $(V ~ , ~ 0 ~ , ~ 0 ~ , ~ 0)$ in
the LVRF characterizing an apparent spontaneous breaking of Lorentz symmetry (SLSB)
and $V$ any constant. 
We can then perform a Lorentz transformation with boost parameter $\Gamma $ from
the LVRF to the rest frame of a UHEP obeying (3), by
writing:
\equation
p_{0} ~ + ~ p_3 ~  ~ = ~ ~ \Gamma ~ ~ (p'_{0} ~ + ~ p'_3)
\endequation
\equation
p_{0} ~ - ~ p_3 ~  ~ = ~ ~ \Gamma ^{-1} ~ ~ (p'_{0} ~ - ~ p'_3)
\endequation
\equation
V ~  ~ = ~ ~ \Gamma ~ ~ (V'_{0} ~ + ~ V'_3) ~  ~ = ~ ~ \Gamma ^{-1} ~ ~ (V'_{0} ~
 - ~ V'_3)
\endequation
\noindent
where the $p_{\mu }$ quadrivector stands now for energy and momentum as measured
in the LVRF, and $p'_{\mu }$ is the same quadrivector as
measured in the rest frame of a UHEP of momentum $p_{\Gamma }$ pointing in
the direction of the spatial axis $i ~ = ~ 3$~. Similar conventions hold for $V_{
\mu }$ and $V'_{\mu }$ . 
Calculations show then that the equations of motion, as seen
in the UHEP rest frame, present a singularity at $b ~ p_{\Gamma }^4 ~ = ~ m^2$~,
i.e. when the deformation term becomes equal to the mass term (well below Planck
scale). 
More details, using explicitly the operator formalism, can be found in references 
[17] and [18], as well as in subsequent
papers of the same series ({\it Deformed Lorentz Symmetry and High-Energy
Astrophysics}, see arXiv.org~).

\section{References}

\re
1.\ Amelino-Camelia, G.\ 2002a, paper gr-qc/0210063 of arXiv.org
\re
2.\ Amelino-Camelia, G.\ 2002b, paper gr-qc/0207049 of arXiv.org
\re
3.\ Chechin V.A., Vavilov Yu.N.\ 1999, Proc. ICRC 1999, paper HE.2.3.07
\re
4.\ Gonzalez-Mestres L.\ 1995, paper hep-ph/9505117
of arXiv.org
\re
5.\ Gonzalez-Mestres L.\ 1996, paper hep-ph/9610474 of arXiv.org
\re
6.\ Gonzalez-Mestres L.\ 1997a, paper physics/9702026 of arXiv.org
\re
7.\ Gonzalez-Mestres L.\ 1997b, paper physics/9704017 of arXiv.org
\re
8.\ Gonzalez-Mestres L.\ 1997c, paper physics/9705031 of arXiv.org
\re
9.\ Gonzalez-Mestres L.\ 1997d, paper nucl-th/9708028 of arXiv.org
\re
10.\ Gonzalez-Mestres L.\ 1997e, paper physics/9709006 of arXiv.org
\re
11.\ Gonzalez-Mestres, L.\ 1997f, paper physics/9712047 of arXiv.org
\re
12.\ Gonzalez-Mestres, L.\ 1997g, paper physics/9712056 of arXiv.org 
\re
13.\ Gonzalez-Mestres L.\ 1999, paper hep-ph/9905430 of arXiv.org
\re
14.\ Gonzalez-Mestres L.\ 2000a, paper physics/0003080 of arXiv.org
\re
15.\ Gonzalez-Mestres L.\ 2000b, paper astro-ph/0011181 of arXiv.org
\re
16.\ Gonzalez-Mestres L.\ 2000c, paper astro-ph/0011182 of arXiv.org
\re
17.\ Gonzalez-Mestres L.\ 2002a, paper hep-th/0208064 of arXiv.org
\re
18.\ Gonzalez-Mestres L.\ 2002b, paper hep-th/0210141 of arXiv.org
\re
19.\ Kirzhnits D.A., Chechin V.A.\ 1971, ZhETF Pis. Red. 4, 261
\re
20.\ Kirzhnits D.A., Chechin V.A.\ 1972, Yad. Fiz. 15, 1051
\re
21.\ Magueijo J. , Smolin L.\ 2001, paper hep-th/0112090 of arXiv.org
\re
22.\ Magueijo, J. and Smolin, L., 2002, paper gr-qc/0207085 of arXiv.org 
\re
23.\ Sato H., Tati T.\ 1972, Progr. Theor. Phys. 47, 1788

\endofpaper
\end{document}